\documentstyle[floats,prl,aps,preprint,tighten,psfig]{revtex} 
 
\begin{document}       
       
\draft       
       
\title{Spiral Surface Growth Without Desorption}  
       
\author{Alain Karma and Mathis Plapp}       
       
\address{       
Department of Physics and Center for Interdisciplinary       
Research on Complex Systems,\\ Northeastern University, Boston,       
MA 02115}       

\date{25 September 1998}

\maketitle       
       
\begin{abstract}
Spiral surface growth is well understood
in the limit where the step motion is
controlled by the local supersaturation  
of adatoms near the spiral ridge. In epitaxial thin-film growth, however, 
spirals can form in a step-flow regime where 
desorption of adatoms is negligible and the ridge dynamics 
is governed by the non-local diffusion field of adatoms on the 
whole surface. We investigate this limit numerically using 
a phase-field formulation of the Burton-Cabrera-Frank model,
as well as analytically. Quantitative predictions, which differ
strikingly from those of the local limit, are made
for the selected step spacing as a function of the 
deposition flux, as well as for the dependence of the
relaxation time to steady-state growth on the
screw dislocation density.
\end{abstract}


\noindent 
Spiral surface growth is one of the most widespread growth 
mechanisms for crystals with atomically flat surfaces. Such 
crystals grow by the incorporation of new atoms at monoatomic 
steps. If steps are pinned at a screw dislocation, they wind 
around the dislocation and form growth spirals.  
The step dynamics, and the final steady-state spacing $l$ between 
successive steps (or equivalently the surface slope $a/l$, 
where $a$ is the lattice parameter) is determined by the 
interplay of surface diffusion, the attachment kinetics 
of atoms at the steps, and the step line tension. 
Recently, there has been renewed 
interest in spiral surface growth
following the observations of spiral ridges in sputtered  
high-temperature superconducting thin films \cite{YBaCuO} 
and in certain semiconductor materials grown 
by molecular beam epitaxy (MBE) \cite{PbTe}.
 
In the classical Burton-Cabrera-Frank 
(BCF) model of surface growth \cite{BCF},
atoms are first adsorbed to the crystalline surface (``adatoms'') and 
then diffuse along the surface until they are either incorporated 
into the crystal at a step, or desorb from the surface with
a probability $1/\tau_s$ per unit time.
Therefore, two different growth regimes can
be distinguished depending on the ratio of $l$ and the
diffusion length $x_s=\sqrt{D\tau_s}$, where $D$ is
the surface diffusion constant. When 
desorption is fast ($x_s\ll l$), 
only adatoms which are deposited near a 
step are incorporated, and the dynamics of the steps is 
{\em local}, that is the velocity of a step is completely 
determined by the local supersaturation and the step curvature. 
This is the well understood regime described by the BCF theory
of spiral growth \cite{BCF,CL}.  
In many practical applications such as MBE, however, 
spirals can form in a step-flow regime at temperatures 
where desorption is negligible. 
Then, all deposited atoms reach 
a step, and successive turns of the spiral are strongly 
coupled via diffusion. The step dynamics becomes a highly non-local 
free boundary problem. The case of steady-state growth
has been investigated by approximate theories
\cite{CC,SHP} and by the boundary integral (potential
theory) method \cite{BI}, but no dynamical solution of
the original equations has been performed to validate
these results.
Another important difference between the two regimes 
is their approach to steady-state growth: in the local limit, 
the spiral finds its final step spacing essentially after a 
single rotation. In contrast, without desorption 
one would expect a slower relaxation to steady-state 
due to the global redistribution of adatoms. In particular,
this relaxation may depend on the 
density of screw dislocations. 
 
The main goal of the present letter is to use 
a phase-field approach to study
the dynamics of spiral ridge formation in the BCF model. 
This approach has recently been used
to efficiently solve a similar free boundary
problem for dendritic growth \cite{KR},
and the mathematical results
of this study are exploited here.
One distinguishing feature of
our approach is that it makes it possible to investigate 
the full crossover from the local to the desorption-free limit,
whereas previous works \cite{FBLH,ABDV} have 
assumed a constant effective supersaturation, as 
appropriate in the local limit. The phase field method
in this context can also be interpreted as a direct continuum
analog to microscopic growth models studied by Monte
Carlo techniques \cite{MC}. 
We make quantitative predictions for the selected step spacing
as a function of the deposition flux
and for the time to approach steady-state growth,
under the assumption that the elastic interaction between
steps can be neglected. We focus mainly
on the situation where adatoms feel the 
same barrier for attaching at ascending or
descending steps, i.e. no Ehrlich-Schwoebel (ES) barrier \cite{ES}, 
in which case attachment can be described 
by a single sticking coefficient $k$.
 
We write the BCF equations in terms of
the dimensionless diffusion field  
$u=\Omega(c-c_{eq}^0)$, 
where $c$ is the adatom concentration, $\Omega$ is the 
atomic area of solid, and $c_{eq}^0$ is the equilibrium 
concentration at a straight step.
The basic equations have the form
\begin{eqnarray} 
& &\frac{\partial u}{\partial t}=D\,\nabla^2 u\,-\,\frac{u}{\tau_s}\, 
+F, \label{e1}\\ 
& &v_n= D\left[\left(\frac{\partial u}{\partial n}\right)_+ 
-\left(\frac{\partial u}{\partial n}\right)_-\right],\label{e2}\\ 
& &u=d_0\kappa, \label{e3} 
\end{eqnarray} 
where $v_n$ is the step normal velocity, 
$(\partial u/\partial n)_{\pm}$ is  
the normal concentration gradient  
on the lower $(+)$ and upper $(-)$ side of the step, 
$\kappa$ is the local step curvature,  
$d_0=\Omega^2c_{eq}^0\gamma/(k_BT)$ 
where $\gamma$ is the step stiffness.
The effective deposition frequency $F$
is related to the actual deposition
flux per atomic area $F_d$ by 
$F = F_d - c_{eq}\Omega/\tau_s$.

We simulate Eqs. 1-3 by reformulating them
in terms of a phase-field model similar to the one used
previously by Liu and Metiu \cite{LM} for a one-dimensional
step train. The basic equations of our model are 
\begin{eqnarray} 
\tau_\psi \frac{\partial \psi}{\partial 
t}&=&-\frac{\delta H}{\delta \psi}~=~W^2\nabla^2\psi\,+\, 
\sin\left(\pi\left[\psi-\psi_s\right]\right)\nonumber\\ 
& &~~~~~~~~~~~~~~\mbox{}+\lambda \,u\left[ 
1+\cos(\pi\left[\psi-\psi_s\right])\right],\label{p1}\\ 
\frac{\partial u}{\partial t}&=&D\nabla^2 u-\frac{u}{\tau_s}\,+F\, 
-\frac{1}{2} \frac{\partial \psi}{\partial t},\label{p2} 
\end{eqnarray} 
where $H$ is the free energy functional depending on the
fields $\psi$ and $u$, $\psi/2$ represents 
the surface height in units of $a$, $\psi_s/2$ is the
height of the initial substrate surface,
$u$ is the concentration field defined 
above, $W$ is the width of the step, $\tau_\psi$
is the characteristic 
time of attachment of adatoms at the steps,
which is typically much smaller than $\tau_s$, and 
$\lambda$ is a dimensionless coupling constant. 
If we replace the coupling between the 
two fields in Eq. \ref{p1} by a constant supersaturation, 
we obtain the continuum limit of the
solid-on-solid model. 
The main difference with the model
of Ref.~\cite{LM} is that the
term $\left[1+\cos(\pi\left[\psi-\psi_s\right])\right]$
is introduced in Eq. \ref{p1} 
to keep the minima of $H$ at fixed values 
($\psi-\psi_s=2n+1$), independently of the adatom concentration. 
A screw dislocation is introduced at the origin by choosing 
$\pi\psi_s$ equal to the polar angle in the $x$-$y$ plane, 
or $\pi\psi_s={\rm atan}(y/x)$, which corresponds to  
shifting the energy minima up by one atomic spacing after one 
complete counter-clockwise rotation. 
 
We now use the recent asymptotic analysis of Karma 
and Rappel \cite{KR} to relate the equations  
of the phase-field model 
to the sharp-interface equations of the BCF model. 
Since the analysis of Ref. \cite{KR} applies directly to 
Eqs. \ref{p1} and \ref{p2}, it need not be 
repeated here. The results of interest are that in the 
thin-interface limit $W/x_s\rightarrow 0$, the phase-field 
equations reduce to the free boundary problem defined 
by Eqs. \ref{e1} and \ref{e2}, with Eq. 3 replaced by 
\begin{equation}
u=d_0\kappa+v_n/k\label{new3}
\end{equation}
where $d_0=a_1W/\lambda$ and $1/k=a_1(\tau_\psi/(\lambda W)-a_2W/D)$. 
The kinetic coefficient $k$ 
can then be made effectively infinite (instantaneous attachment), 
thereby recovering the condition (\ref{e3}) 
of local equilibrium, by choosing the coupling constant 
$\lambda=\tau_\psi D/(a_2W^2)$. 
The numerical constants $a_1$ and $a_2$ are  
determined by the form of the energy function in Eq. \ref{p1} 
and can be evaluated using Eqs. 51, 54-56, and 58-59 in 
Ref. \cite{KR}, together with the one-dimensional  
stationary profile solution of Eq. \ref{p1} with $u=0$ for 
an isolated step with $\psi(\mp \infty)=\pm 1$, 
\begin{equation} 
\psi_0(x)=1-\frac{4}{\pi} {\rm 
atan}\left(\exp\frac{\sqrt{\pi}x}{W}\right). 
\end{equation} 
The resulting numerical values are  
$a_1=0.718348$ and $a_2=0.510442$. 
Eqs. \ref{p1} and \ref{p2} were simulated on 
a square lattice of edge length $L=N\Delta x$ with 
zero flux boundary conditions and $N$ varying between
100 and 600. 
As in Ref. \cite{KR}, Eq. \ref{p1} was integrated 
using an explicit Euler scheme, Eq. \ref{p2}  
using a Crank-Nicholson implicit 
scheme. For the simulations, we measured lengths and
times in units of the phase-field parameters,
i.~e. $W = \tau_\psi = 1$. In these units, we chose
$D=10$, which yields $\lambda=19.591$,
$d_0=0.0366$, $\Delta x=0.5$, and $\Delta t=0.025$.
The final results, stated in dimensionless ratios
of physically well-defined quantities, do not depend
on the choice of $W$ and $\tau_\psi$ as long as
$W\ll x_s$, $W\ll l$, and $\tau_\psi \ll 1/F$.

\begin{figure} 
\centerline{\psfig{file=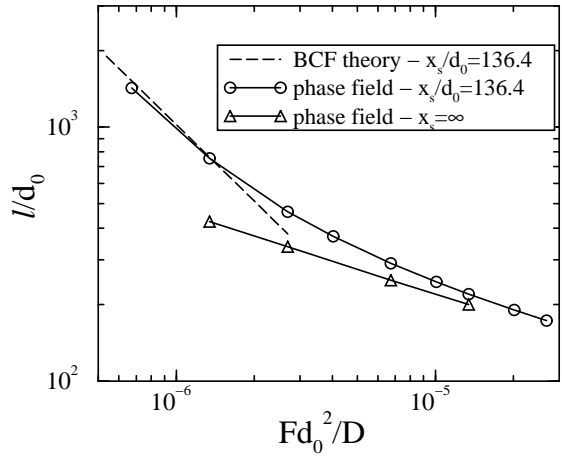,width=.5\textwidth}}
\vspace{0.3cm} 
\caption{Step spacing $l$  
normalized by $d_0$ as a function of 
$F$ for finite and vanishing desorption. 
The standard BCF theory in the local limit is shown as
a dashed line for comparison.} 
\label{scaling} 
\end{figure} 
Simulations were started from a straight ridge along 
the $x$-axis with one end pinned to the screw dislocation  
($\psi=\psi_s$). After a long transient, the 
ridge reaches a steady-state spiral shape  
with a constant angular rotation frequency $\omega$, 
and a step spacing $l$. For our plots, we defined $l$
to be the distance between the first two successive steps from the 
center. This value is only a few percent smaller than the asymptotic 
step spacing far from the core. A plot of $l$ 
as a function of $F$ is shown in Fig. \ref{scaling} for  
$x_s/d_0=136.4$, and $x_s=\infty$ (no desorption). 
 
Let us now compare our results to the standard
BCF theory of spiral growth that is based on the
expression for the normal step velocity 
\begin{equation} 
v_n=v_\infty\left(1-r_c\kappa\right) 
\label{local} 
\end{equation} 
where $v_\infty$ is the velocity of a straight step 
and $r_c$ is the critical radius for island nucleation. 
Expressions for $\omega$ and $l$ are then obtained 
by looking for shape preserving solutions of Eq. \ref{local} 
in a frame rotating at angular frequency $\omega$. 
In the local limit, $l\gg x_s$, the solution is \cite{CL}:
\begin{eqnarray} 
l&=&19\,r_c\label{bcfl}\\ 
\omega &=&\frac{2\pi}{19}~\frac{v_\infty}{r_c}\label{bcfo}, 
\end{eqnarray} 
with $v_\infty=2Fx_s$ and $r_c=d_0/(F\tau_s)$;  
$F\tau_s$ is the supersaturation far away from steps. 
The corresponding curve for $l$ 
is plotted as a dashed line in Fig. \ref{scaling}. 
For an arbitrary ratio $l/x_s$, Eq. \ref{local} is 
no longer exact because the diffusion fields of
the steps overlap.
Cabrera and Coleman (CC) proposed \cite{CC} that 
a rough estimate of $l$ and $\omega$ can 
be obtained by assuming that  
Eqs. \ref{bcfl}-\ref{bcfo} continue to hold with  
\begin{eqnarray} 
v_\infty&=&2Fx_s\tanh (l/2x_s)\label{cc1},\\ 
r_c&=&d_0/u(0)=d_0/\left[F\tau_s\left(1-1/I_0(l/x_s)\right)\right], 
\end{eqnarray} 
where Eq. \ref{cc1} 
is the exact expression for the velocity of an infinite  
one-dimensional step train of spacing $l$, and 
$u(0)$ is the supersaturation 
at the center of a circular terrace of radius $l$,
obtained by solving Eq. \ref{e1} subject 
to the boundary condition $u(l)=0$. This yields 
$u(r)=F\tau_s[1-I_0(r/x_s)/I_0(l/x_s)]$, 
where $I_0$ is the zero-th order  
modified Bessel function. 
One interesting consequence of the CC estimate 
is that in the limit $x_s/l\rightarrow \infty$ 
\begin{eqnarray} 
l &=& A(d_0D/F)^{1/3}\label{cca1}\\ 
\omega &=& 2\pi F\label{cca2} 
\end{eqnarray} 
where $A=76^{1/3}\approx 4.236$. The expression for $\omega$ is 
exact and follows from global mass conservation: 
since for $l\ll x_s$ all adatoms reach a step, the spiral 
rotation frequency must just be $F$ in steady-state. 
The same scaling relations, but with a different
prefactor $A$ have been obtained by the boundary 
integral method \cite{BI}. The 
scaling for $l$ becomes exact 
in the limit $d_0/l\rightarrow 0$, which is practically
always satisfied.
This can be understood from 
a simple dimensional analysis. In the limit 
where $\tau_s\rightarrow \infty$, all parameters 
can be removed from the BCF equations by making the 
variable transformations  
\begin{equation} 
t'=tF,~ x'=x(F/d_0D)^{1/3},~ 
u'=u(D/d_0^2F)^{1/3},\label{scale} 
\end{equation} 
if one neglects $\partial_t u$ in Eq. 1 which is
of relative magnitude $d_0/l$.
Our numerical results confirm this scaling and yield 
a value of $A=4.626$. They also show that
the cross-over from $l\sim F^{-1}$ to $l\sim F^{-1/3}$
is very slow.
 
\begin{figure} 
\centerline{\psfig{file=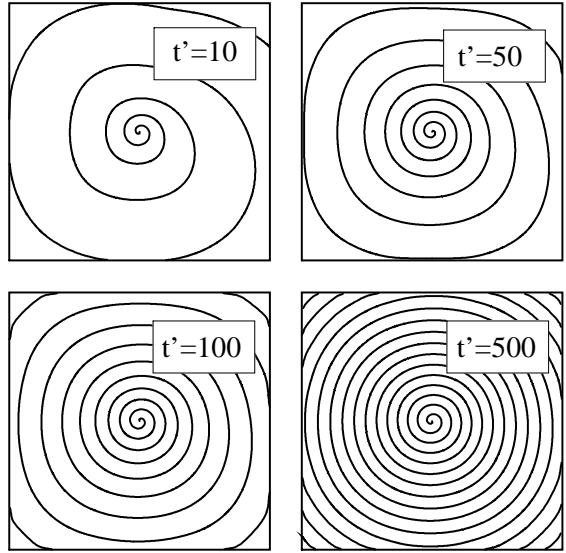,width=.6\textwidth}} 
\vspace{0.3cm} 
\caption{Spiral ridge at different
times $t'=tF$, equivalent to the 
number of monolayers deposited, for $x_s=\infty$, 
$Fd_0^2/D=1.344\times10^{-5}$,
and $L/d_0=5464$.} 
\label{shapes} 
\end{figure}
\begin{figure} 
\centerline{\psfig{file=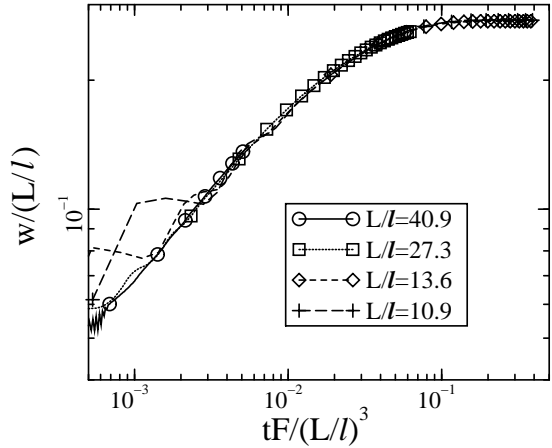,width=.5\textwidth}} 
\vspace{0.3cm}
\caption{Rescaled surface width $w/(L/l)$ as a function of 
rescaled time $tF/(L/l)^3$ for four different system sizes. 
The large fluctuations at early times occur during the initial
winding of the straight ridge into a spiral shape.}
\label{wfig} 
\end{figure}

 
Next, let us examine how the dynamical aspects of
the spiral ridge formation depend on the system size $L$,
which on a real surface is roughly given 
by the mean distance between dislocations.
In the local limit $x_s\ll l$,  
$\omega$ and $l$ are sharply selected
on a time scale of one rotation, and the ridge 
winds itself into a classic Archimedian spiral 
(i.e. spiral with a constant $l$) in $\sim L/l$
rotations. In contrast, Fig. \ref{shapes} shows 
that, in the desorption-free limit, the transient spiral ridge 
evolves extremely slowly towards an Archimedian 
spiral via a progressive reduction of the
step spacing away from the core. The 
rotation frequency of the center is faster than $F$
at the beginning, as a larger step spacing allows more
adatoms to attach to the ridge, and then
slowly approaches $F$. In addition, Fig. \ref{shapes} shows
that the shape is strongly influenced by the 
boundaries during the initial transient.
The final spacing, however, is independent of the system size.
In order to quantify 
the transient dynamics, 
we calculate the surface width  
\begin{equation} 
w(t)=\frac{1}{2}<\psi(x,t)^2-<\psi(x,t)>^2>^{1/2} 
\end{equation} 
where $<f>\equiv L^{-2}\int f dxdy$.
In steady-state, we simply have
$w(t\rightarrow \infty)\propto L/l$ for $L\gg l$. 
We plot in Fig. \ref{wfig} $w(t)/(L/l)$ 
as a function of $Ft/(L/l)^3$
for different system sizes.
The data collapse remarkably onto a single curve, which shows 
that the time to reach steady state  
scales as $(L/l)^3$. This cubic law can be
understood analytically by considering the train dynamics
in the region away from the spiral core. In this region, 
the effect of the step curvature can be neglected
and a one-dimensional step train is governed by
the simple evolution equation 
$\partial_tl_n = (F/2)(l_{n-1}-l_{n+1})$, 
where $l_n$ is the distance between the $n$th and $n+1$th steps. 
This set of discrete equations can be transformed into 
a continuum equation for the coarse-grained surface 
height $h(x,t)$ (in units of $a$) of 
the standard conserved form \cite{V}: 
\begin{equation} 
\partial_t h = F - \partial_x J,
\label{heq1} 
\end{equation} 
where $J$ is the surface current. Here,
$J=-D\partial_x <u>$ where
$<u>=Fl^2/12D=F/12D(\partial_x h)^2$ is the
the average adatom concentration between two steps,
which combined with Eq. \ref{heq1} yields 
\begin{equation} 
\partial_t h = F + \frac{F}{12}\,\partial_x^2 
\left(\frac{1}{(\partial_x h)^2}\right).
\label{heq2} 
\end{equation} 
This equation has a scaling solution of the form
\begin{equation} 
h(x,t) = Ft + (L/l)\tilde h(x/(L/l), t F/(L/l)^3),
\end{equation} 
where the system size drops out of the resulting equation for 
$\tilde h$. Thus, the relaxation time depends on the 
third power of the system size as observed in
our simulations. 
 
Let us now examine how these results 
become modified when the assumption of local 
equilibrium at the step is relaxed by letting $k$
be finite in Eq. \ref{new3}.
In this case, an approximate expression for $l$ 
can be obtained by repeating 
the estimate of CC with the modified boundary condition 
$u(l)=v_\infty/k=Fl/k$ at the edge of the  
circular terrace around the dislocation center. 
This yields $u(0)=(1+4D/l k)Fl^2/4D$, where  
as before $l=19d_0/u(0)$. So for fast attachment
($4D/l k\ll 1$) we recover the previous scaling 
$l\sim F^{-1/3}$, whereas for slow attachment ($4D/l k\gg 1$)
we obtain the scaling 
$l\sim (d_0k/F)^{1/2}$
in agreement with boundary integral results \cite{BI}.
We performed a series of simulations
with a finite $k$ by choosing $\lambda=1$ and observed
a slow cross-over towards $F^{-1/2}$ 
with increasing $F$, which is consistent 
with this prediction. The spiral ridge was
also found to relax much slower than
in the local limit but we did not conduct a systematic finite
size scaling analysis to determine the dependence on $L$. Finally,
if the present analyses are extended to the case of a finite
ES barrier, one concludes that the scaling $l\sim F^{-1/3}$
remains unchanged. In this case, however,
the step train away from the spiral core is known to be
morphologically unstable \cite{BZ}, and this instability
may alter spiral growth in ways that
remain to be investigated.
 
Several of the present predictions should be
experimentally testable. If desorption is negligible,
one should observe a dependence of the form $l\sim F^{-\alpha}$
with $\alpha$ between $1/3$ and $1/2$.
Moreover, by examining the ridge dynamics one
can obtain information about the relative importance of
desorption and diffusion.
In particular, in experiments the distance $L$ between
screw dislocation centers is often in the range of 
five to ten $l$ \cite{YBaCuO,PbTe}. 
If desorption is negligible,
the cubic dependence of the relaxation time
to steady-state growth on $l$ implies that the
spiral ridge should reach a maximum surface slope 
only after a few hundred monolayers are deposited.
 
The present phase-field 
approach should be generally applicable to 
simulate a wide range of mesoscopic surface  
growth phenomena.
Interesting future prospects are to include concentration 
fluctuations to study the crossover from  
spiral growth to island nucleation at high 
temperature/flux, to study the effect of
anisotropy of the line tension and/or the attachment
kinetics on the growth morphology, and 
to include an ES barrier. 
Finally, elastic effects which have recently been shown to
influence spiral growth dynamics \cite{ABDV} could be 
incorporated by coupling the dynamics of $\psi$ 
and $u$ to the strain field.

This research was supported by US DOE
grant No DE-FG02-92ER45471 and benefited
from supercomputer time at NERSC. We thank R. Kohn,
D. Wolf, and A. Zangwill for valuable exchanges.

\end{document}